\documentclass[a4paper]{article}
\usepackage{INTERSPEECH2021}

\title{Models of Music Cognition and Composition}
\name{Abhimanyu Sethia, Aayush}
\address{
  Indian Institute of Technology, Kanpur}
\email{sethia@iitk.ac.in, paayush@iitk.ac.in}

\begin{document}

\maketitle
\begin{abstract}
  Much like most of cognition research, music cognition is an interdisciplinary field, which attempts to apply methods of cognitive science (neurological, computational and experimental) to understand the perception and process of composition of music.
  
  In this paper, we first motivate why music is relevant to cognitive scientists and give an overview of the approaches to computational modelling of music cognition. 
  
  We then review literature on the various models of music perception, including non-computational models, computational non-cognitive models and computational cognitive models. Lastly, we review literature on modelling the creative behaviour and on computer systems capable of composing music. 
  
  Since a lot of technical terms from music theory have been used, we have appended a list of relevant terms and their definitions at the end.
\end{abstract}

\section{Introduction}
Pinker  once famously said “[as far as biological cause and effect are concerned] music is useless.”\cite{pinker} One may wonder why music should be of interest to cognitive scientists. Three arguments have been advanced in this regard \cite{intro}.  First, music has been identified as a universal human trait and plays a crucial role in everyday life. Second, music has a central role in ontogenetic development and human evolution. Characteristics of music have been identified as ‘evolutionary adaptive behaviour.’ \cite{evolutionary} Third, music listening, performance and interaction involve a wide range of cognitive, perceptual and emotional processes, which makes it an ideal object for the study of the human mind.
\subsection{Computational modelling of Music Cognition}

Longuet‐Higgins in 1976, in his pioneering paper, presented a computer program which he claimed, would capture the perception of tonal and rhythmic relationships between notes of Western melodies.\cite{Higgins1976} 

Historically, two contrasting approaches have been followed for computational modelling music perception, namely descriptive modelling and explanatory modelling.\cite{oxford} The former attempts to model what a phenomenon does, while the latter attempt to model how it does it. In our context, a descriptive model aims at modelling mental processes that occur when listening to or composing music, while an explanatory model is based on the formalisation of musical knowledge and theory.

\section{Models of Music Perception}
\subsection{Non-Computational Models}
Simon and Sumner(1968)\cite{[2]} attempted to define a formal language to describe patterns perceived by humans in processing musical sequences. Their language consisted of alphabets (symbols to represent the range of possible values of a particular musical dimension) and three kinds of operations-
\begin{itemize}
    \item Subset operation: to derive more abstract alphabets from the current ones
    \item Operations that relate a symbol to its predecessor, thereby describing a pattern of symbols
    \item Replacing a pattern with an abstract alphabet
\end{itemize}

According to this model, when we listen to music, we induce an alphabet, recognise a pattern and then use the pattern to extrapolate the sequence.

The other notable non-computational theory is the Generative Theory of Tonal Music (GTTM) \cite{[3]}. GTTM is inspired by use of Chomskian grammar to define language. The theory essentially claims that a listener inherently infers 4 types of hierarchical structures in a musical surface. It is beyond the scope of this paper to describe these structures in detail, but it suffices to say that the theory has received support from various experimental studies (\cite{deliege}, \cite{palmer}).

\subsection{Temperley's Model}
Perhaps the most complete computational theory of music cognition, to date is that presented by Temperley in 2001 \cite{[5]}. The theory attempts to explain music cognition by generating structural descriptions from musical ‘surfaces’. However, in 2007, Temperley improved some of these rule-based models into a Bayesian-probability based model, to infer the key of a melody. His model is based on three basic principles of music theory: one, melodies remain in a narrow pitch range; two, intervals between notes in a melody are small and three, depending on the key, notes follow a distribution.

\subsection{Key Recognition Models}
Key is one of the essential elements required to get an understanding of a musical piece. Experimental psychology has shown that people, irrespective of whether they have formal music training or not, have an inherent understanding of Key. Krumhansl in 1990 \cite{[6]} showed that given a tonal context, listeners could adjudge some notes to fit better with the given key than others.

Key Recognition is one of the central areas of research in music cognition, and naturally, it has invited an array of computational model theories as well. Perhaps the most famous among these is the classic work of Longuet-Higgins and Steedman, published in 1971.\cite{[7]}

The Longuet-Higgins Steedman (abbreviated as LH-S) model is a brute force approach, which processes the notes in a melody one by one in a sequential fashion. For every note, the key whose scale does not include that particular note is eliminated. If a unique key remains at any point in time, then that key is selected as the key of the piece. For other cases, the model utilizes some advanced music theory notions that are not relevant from a cognitive science perspective. The LH-S model yielded good results on simpler music melodies but failed miserably on complicated ones. Also, it’s easy to see that the model had little to no neuro-architectural basis, and the brute-force elimination method was merely a computational choice.

The model of Krumhansl and Schmuckler, published in 1990, extends the LH-S model significantly to counter its shortcomings. Rather than focusing explicitly on eliminating possibilities, the extended model tried to measure the likelihood of a melody having a particular key. The model proposed the concept of a \emph{key profile}, i.e., a vector with 12 values where each value was a measure of the compatibility of a note with the given key. These key profiles were obtained using experiments on human subjects, where they were asked to rate how well the notes \emph{fit} with a given key. \cite{[8]}

Given the key profiles, the model processes the melody in the same sequential fashion as the LH-S model and generates another 12-valued \emph{input vector}. This generated vector is then compared with all the key profiles, and the key profile with the maximum similarity is output as the required key.

Krumhansl observed that this type of approach was of a \emph{template fitting} nature. The KS model was considerably efficient and consequently gave rise to a whole new class of models in the field of computational Key Recognition research, known as \emph{Distributional models}. The name was courtesy of the fact that these models inherently relied on the distribution of various notes/pitches in a musical piece while completely disregarding the temporal arrangement of the notes. This class of distributional models essentially followed the same approach as the KS model. The differences mainly lay in the theoretical music concepts used for key profiling and likelihood estimation.

One of the major problems faced by these distributional models was recognizing changing of keys within a piece. Huron and Parncutt’s model\cite{[9]} (1993) solved this by updating the generated input vector by exponentially decaying the contributions of previous notes. This ensured that a change in the key midway during a piece was encapsulated by a change in the input vector, and consequently, the recognized key output.

The modifications in the LH-S and KS models and better insights into human perception of music through experimental psychology have led to computational models that are highly efficient at recognizing the key (or keys) of a musical piece. But this does not hide the fact that this is an oversimplified approach. For example, these models completely disregard the temporal arrangements of notes, but experiments have demonstrated that different arrangements of notes can imply different key perceptions \cite{matsunga}.

\subsection{Melodic Expectation Models}
In music cognition, melodic expectation refers to the deductive cognitive behaviour which takes place in the listener’s brain in response to music. For example, if a listener familiar with Indian Classical music listens to a partial octave “Sa Re Ga Ma Pa Dha Ni”, they’ll have a tendency to expect another “Sa” to follow.

\subsubsection{Meyer’s Model}
Meyer, in his 1956 book Emotion and Meaning in Music \cite{meyer}, discusses the relationship between the cognitive processes which take place when we listen to a musical piece, the expectations they generate, and it’s perceived meaning.

Meyer classifies the perceived meaning of a musical piece into two classes: absolute and referential. Absolute meaning refers to the meaning the piece acquires through different musical elements such as melody, lyrics, etc. Referential meaning refers to the meaning acquired through non-musical elements. For example, a “happy” melody in a song contributes to the absolute meaning of the song, while a personal event which you might associate with the song adds to the referential meaning. Meyer’s model (and all other models based on it) focus on the absolute meaning. 

Henceforth, the term meaning refers to the absolute meaning unless specified otherwise. Meyer proposes that the perceived meaning is directly related to how a listener’s expectations are met or violated. He mentions three ways in which expectations can be violated: delaying the expected consequent; the antecedent generating ambiguous expectations; and the consequent being unexpected.

Building upon his theories of meaning and expectation, Meyer proposes a probabilistic model for generation of expectations based on antecedent musical intervals. He suggests that once a listener gets accustomed to a particular music style, the expectations are embedded as Markov chains in their cognition and they develop an internal understanding of these probabilistic processes.   

One of the problems with Meyer’s model is that it fails to distinguish between perpetual uncertainty (uncertainty caused by the listener’s cognitive expectations) and stylistic uncertainty (uncertainty inherent in the musical piece).

\subsubsection{Implication-realisation theory}
Perhaps the most famous model of melodic cognition, the Implication-realisation theory was developed and published by Narmour\cite{IR} between 1990-1992. It is based on Meyer’s model and combines two independent approaches: a universal bottom-up system based on unconscious and automatic cognitive processes, and a top-down approach accounting for the learned and cultural influences. 

The bottom-up approach is based on what Narmour calls “two universal hypotheses”:
\begin{enumerate}
    \item Hearing two similar items yields an expectation of another similar item.
    \item Hearing two different items yields an expectation of change
\end{enumerate}

Narmour describes the property of closure of a musical interval by listing various factors which contribute to it (such as a rest following the interval, a higher second tone relative to the first tone of an interval, etc). An interval with a strong closure generally signifies the end of a musical structure, while an unclosed interval (called an \emph{implicative interval}) generates expectancies for the subsequent interval (called a \emph{realised interval}).

Narmour goes on to describe the various types of expectancies generated by different implicative intervals. He suggests expectancy patterns by making use of Gestalt principles of proximity, similarity, closure and symmetry. For example, Narmour suggests that a small melodic interval is expected to be followed by another interval of similar size, whereas a larger melodic interval generates an expectancy of a smaller consequent interval (a principle known as \emph{intervallic difference}). Similarly, the principle of registral direction states that a smaller interval implies a continuation of pitch direction (increasing, decreasing or static) while a larger interval implies a reversal.  

The top-down approach of Narmour’s model underlines the influence of musical knowledge, acquired through experience, on melodic expectancy. These influences can be varied in nature, ranging from a knowledge of music theory, style, and structures to a familiarity with particular genres, artists, etc.  

\subsubsection{A statistical model}
Marcus Pearce and Geraint Wiggins (2006)\cite{pearce} argue that Narmour’s approach is too inflexible to account for the influence of context and learning on expectations. According to them, the quantitative approach of IR theory is descriptive but not explanatory, and that it fails to underline the cognitive aspects responsible for the behaviour. 

Instead, they suggest that a statistical approach for expectancy generation, instead of a quantitative and analytical model like I-R, would be able to capture the behavioral data in a much better way. They present an alternative model based on a statistical model, possessing no prior musical knowledge or experience. They propose that an exposure to a large corpus of musical data would help it encode probabilities for different types of melodic expectations.

In their 2006 paper, they use an n-gram model (which is popular in language processing domains) to generate the expectancy probabilities. The musical data is fed in the form of a sequence of events, with each event consisting of some basic features (such as pitch range, onset time and duration).

\subsection{Ritardandi Models}
The final ritardandi, or the ‘final retard’ refers to the typical slowing down at the end of the music performance. We will look at two approaches used to model the expressive timing of the ritardandi. 
\subsubsection{Kinematics Model}
The kinematic models \cite{kinematics} draw a parallel between a music performances and physical motion in the real world. This is based on the research that musicians, in using tempo and timing as an expressive device, allude to physical motion. 

In context of the final ritard, the analogous case is the deceleration in human motion, which can be summarised by the often used equation $v(t)=u+at$.

But here, u is the initial tempo, a is the acceleration factor (will be deceleration in this case) and v is the tempo as a function of time (as it slows down). The equation is then generalised and expressed in terms of the score position. This gives a model of the tempo (relative to the initial or pre-ritardando tempo)  as a function of the score position (normalised w.r.t to the length of the ritard). The model was observed to fit well with the empirical data (\cite{[10]}). 

\subsubsection{Perception Based Model}
This model \cite{perception} consists of two components. First, a model of perceived regularity (or tempo tracker), which tracks the perceived tempo of a performance. Second, is a model of rhythmic categorisation (or quantizer). It take the residue from the first model, that is the timing pattern after a tempo interpretation, and predicts the perceived duration category. 

A composer/performer would want the listener to recognise the original rhythm. The model makes predictions on if a rhythm performed with tempo and timing variations, would still be recognisable to the listener and when the perceived rhythmic structure would break down due to too much tempo change. Hence, it predicts the perceptual boundaries between which a ritardandi is supposed to occur.

Both models can be seen as tempo predictor.

\section{Models of Music Composition}
\subsection{Modelling Creative Behaviour}
Boden \cite{[11]} modelled creative behaviour using the notion of a conceptual space and its exploration by creative agents. He defined a conceptual space as the set of acceptable concepts, exploratory creativity as the process of exploring a given conceptual space and transformational creativity as the process of changing the rules that define the conceptual space.

Various other computational cognitive models of creativity exist but are not detailed enough for implementation (\cite{wallas}, \cite{koestler}).

Wiggins \cite{[12]} formalised Boden’s model through his ‘creative systems framework (CSF)’. CSF consists of :
\begin{itemize}
    \item A mutable rule set R which defines the conceptual space
    \item A mutable rule set E, which evaluates the quality of the items created
    \item And a traversal strategy T, which is used by a creative agent to explore the conceptual space
\end{itemize}
Note that R and E are different notions. For example, one may recognise a joke, yet acknowledge that it is not a good one.

\subsection{Musical Composition Systems}
We first look at the non-cognitive systems. In most of the non-cognitive musical systems, the emphasis is on reproducing the existing composers and their styles. CHORAL\cite{choral} (1988), is one such rule-based expert system implemented in a special language (called BSL). It is used for harmonization of chorale melodies in the style of JS Bach. HERMAN \cite{herman} (1998) is another such system, which generates music and allows user to vary emotional property from neutral to scary. It must be noted that these models are descriptive and not explanatory, i.e. they rely on music theory and encode it. \cite{oxford} argues that a system must contain significant amount of autonomous learning for it to qualify as genuinely ‘creative,’ which the non-cognitive musical composition systems lack. 

Very few computational cognitive models of music composition exist, most notable of which is \cite{conklin}. In this paper, Conklin presented four method of generating music from a statistical model, one of which is sequential random sampling of events. However, it generates events in a random walk and so is susceptible to straying into a local minima in the space of possible compositions.  This shortcoming was addressed by the Hidden Markov Model (HMM) \cite{rabiner}, which generates events from hidden states.

Allan and Williams proposed another such Hidden Markov Model for music composition in \cite{hmm}. Given a melody, the harmonisation model attempts to create three further lines of music which will sound pleasant when played in harmonic superposition with the original melody. In this Hidden Markov Model, the visible states are melody notes and a sequence of observed events make up the melody line. While the hidden states are chords and a sequence of hidden events make up a possible harmonisation for a melody line. Here, \cite{hmm} models how a visible melody line is emitted by a hidden sequence of harmonies.

\bibliographystyle{IEEEtran}
\bibliography{mybib}

\section{Appendix: Terms from Music Theory}
\subsection{Pitch}
Pitch is a property of sound, characterized by the frequency of the sound emitted, which makes it possible for humans to perceive sounds as being relatively “higher” or “lower”. Pitch can be determined only if the frequency is clear and stable enough. Noises in the frequency make it difficult for beings to perceive the pitch of the sound.

\subsection{Notes}
A note is the most basic unit in music theory. There are essentially only 12 notes in music, 7 fundamental and 5 derived (called accidentals, characterized by a \# (sharp) or a b (flat)).  The 12 notes are C, C\#, D, D\#, E, F, F\#, G, G\#, A, A\#, B. An accidental note is midway between two fundamental notes. For example, C\# has the average frequency of C and D. In musical terms, C\# is a semitone higher than C or a semitone lower than D. C\# can also be denoted as Db, and the two notations are used interchangeably. 

The best way to understand notes in music theory is to think of an analogy to the colour spectrum. The colour spectrum encompasses all the frequencies within the visible range, which implies an infinite number of frequencies. But the spectrum is mainly perceived as seven colours. In music theory, the pitch is analogous to the frequency, while a note is analogous to a colour. Although we’ll use the terms pitch and notes interchangeably in this paper, it’s important to keep the subtle difference in mind.

A continuous linear sequence of musical notes in time is called a melody. The continuous characteristic here refers to the perception, i.e. a listener perceives the sequence as a single entity.
\subsection{Octaves}

Humans’ perception of notes (or pitch, in general) has a very special characteristic, which is termed as “the basic miracle of music”: if pitch X has double the frequency of pitch Y, then the brain perceives them as being similar. This gives rise to octaves, another fundamental element in music theory.

An octave is defined as the interval between one note and another note with double its frequency (Sa, Re, Ga, Ma, Pa, Dha, Ni, Sa constitutes 1 octave, ending on twice the initial frequency. Although the last Sa is higher than the first Sa, we perceive them as belonging to a similar class). This repetitiveness is elementary to music perception and production and gives rise to the concept of keys and scales.

\subsection{Pitch Notation}
A pitch is denoted by combining a note notation with a number. The note notation refers to the pitch class it belongs to, while the number is representative of the octave of the pitch. The frequency of ~16.35 Hz is used as a reference point and is assigned the label C0. So, C1 is 32.70 Hz, C2 is 65.41 Hz, C3 is 130.81 and so on.

\subsection{Scale}
A scale is an ordering of notes in an ascending or descending order of frequencies. Most scales are repetitive; that is, the pattern of notes is the same for every octave (example, C3 D3 E3 F3 G3 A4 B4 C4 D4 E4 F4… is a scale with seven notes per octave). Scales are classified into different types depending on the number of notes they contain per octave, e.g. chromatic (12 notes per octave), octatonic (8 notes per octave, used in modern classical music), heptatonic (7 notes per octave, popular in Western Music), etc. 

\subsection{Keys and Chords}
In simple terms, a key refers to a group of pitches/notes (generally seven in number) that occur together and form the basis of classical and western music. For example, the key of C consists of the notes C, D, E, F, G, A, B\; the key of D consists of notes D, E, F\#, G, A, B, and C\#. 

A chord is a harmonic superposition of multiple notes (usually greater than or equal to three in number). Chords are one of the essential elements of music composition and are typically produced by combining notes belonging to the same key. 

A musical piece is generally (but not necessarily) and majorly written using the notes and chords of one key, and that key is referred to as “the key” of that piece. Changes in key or slight modulations are commonplace in musical compositions. 

\subsection{Temporal Units in music}
\begin{itemize}
    \item A \textbf{beat} is the standard unit of measuring time in music. A given number of beats form a measure or a bar. 
    
    The value of one beat is not pre-defined and is expressed in terms of the “fraction of a note.” Also, the number of beats in a measure changes across different music pieces. The time signature of a song determines these two entities. The time signature is denoted by a fraction. The numerator defines the number of beats in a measure, and the lower number indicates what fraction value is considered equal to a beat.
    
    For example, the classic “Clap Clap Stomp” intro to “We Will Rock You” is in a 4/4 time signature. This means that each measure has four beats, and each beat is equal to a quarter of a note. The clap is counted as a quarter note (1 beat), while the stomp is counted as a half note (2 beats), making a total of 4 beats. Hence, two claps and a stomp form one measure.  
    
    \item \textbf{Tempo} refers to the overall speed of the song and is expressed in terms of “beats per minute (BPM).” A higher BPM generally indicates a faster piece.

    \item \textbf{Meter}: An overly simplified definition of meter is that it refers to the upper number in a time signature. Broadly speaking, meter refers to the underlying property of music which allows  to divide and group the beats. Continuing with our previous example, meter would refer to the grouping of “two claps and one stomp” together.
    
    \item \textbf{Rhythm} is a loose term used to indicate the general temporal structure of a musical piece, and to show how music moves through time. Rhythm is very much dependent on the concepts of tempo and metre, as both of them dictate the temporal arrangement and movement of a music piece.
    
    \item \textbf{Ritardando} refers to a gradual slowing down of tempo within a piece of music. A prominent example of this effect can be found towards the ending of many classical and Western songs, where the performer tends to decrease the tempo before ending the song, somewhat analogous to how a car driver tends to slow down the car gradually before stopping it completely. 

\end{itemize}

\end{document}